\documentclass[aps,prb,twocolumn,amsmath,amssymb,bibnotes]{revtex4}

\usepackage{color,epsfig,rotating,graphicx,subfigure,comment}

\usepackage{amsmath,amssymb,amscd,ulem}

\usepackage[latin1]{inputenc}
\usepackage[english]{babel}

\usepackage{dsfont}
\renewcommand{\Re}{{\rm Re}}
\renewcommand{\Im}{{\rm Im}}

\newcommand{\ri}{{\rm i}}
\newcommand{\re}{{\rm e}}
\newcommand{\rd}{{\rm d}}

\newcommand{\Tr}{{\rm Tr}}

\begin{document}
\title{Radiative thermal switch exploiting hyperbolic surface phonon polaritons}

\date{\today}

\author{Annika Ott$^1$, Yang Hu$^{2,3,4}$ , Xiao-Hu Wu$^4$, Svend-Age Biehs$^{1}$}
\email{ s.age.biehs@uni-oldenburg.de} 
\affiliation{$^1$ Institut f\"{u}r Physik, Carl von Ossietzky Universit\"{a}t, D-26111 Oldenburg, Germany}
\affiliation{$^2$ Basic Research Center, School of Power and Energy, Northwestern Polytechnical University, Xi'an, 710064, P.R. China}
\affiliation{$^3$ Center of Computational Physics and Energy Science, Yangtze River Delta Research Institute of NPU, Northwestern Polytechnical University, Taicang, Jiangsu, 215400, P.R. China}
\affiliation{$^4$ Shandong Institute of Advanced Technology, Jinan 250100, P.R. China}

\begin{abstract}  
We study the radiative heat flux between two nanoparticles in close vicinity to the natural hyperbolic material hBN with its optical axis oriented parallel to the interface. We show that the heat flux between the nanoparticles can be efficiently modulated when rotating the nanoparticles due to the coupling to the highly directional hyperbolic surface modes in hBN. Finally, we discuss the thickness and distance dependence of this effect. 
\end{abstract}

\maketitle

%
%
\section{Introduction}

Recently, several possibilities for the manipulation of near-field radiative heat transport due to many-body interactions have been highlighted:
It could be shown that the radiative heat flux (HF) between two objects can be tremendously enhanced by the interaction with an environment
which has interfaces supporting long-range surface modes~\cite{Saaskilathi2014,Asheichyk2017,DongEtAl2018,paper_2sic,HeEtAl2019}. Such an 
enhancement can also be observed for the HF through plasmonic or hyperbolic films~\cite{MessinaEtAl2012,MessinaEtAl2016,ZhangEtAl2019b,McSherryLenert2020}  due to the transmission of large wave-vector modes through such structures. Recently, non-reciprocal media such as magneto-optical materials or Weyl metals have been studied in many-body assemblies because they exhibit, on the one hand, interesting effects like a persistent heat current~\cite{zhufan,zhufan2}, giant magneto-resistance~\cite{Latella2017,Cuevas,HeEtAl2020, Cuevas2,Song}, a Hall and anomalous Hall effect for thermal radiation~\cite{hall,Ahall} as well as a circular near-field HF which is closely connected spin and angular momenta of thermal radiation~\cite{Ott2018,Zubin2019,Khandekar}. On the other hand, such non-reciprocal media allow for actively controlling the directionality of surface modes and therefore can be used to control the strength of HF between two objects via the non-reciprocal surface modes of the environment which can result in a strong HF rectification~\cite{Ott2019,Ott2020}. Another possibility is to use the intrinsic anisotropy of the environment as encountered in uni-axial materials as shown for the HF between two nanoparticles (NPs) in the vicinity of a 2D meta-surface made of graphene strips~\cite{ZhangEtAl2019}; graphene also offers the possibility for active thermal switching by electrical biasing~\cite{YuEtAl2017,IlicEtAl2018} which is much more efficient than electrical switching with ferroelectric materials~\cite{HuangEtAl2014}. A review of these effects and other recent developments in the field of many-body effects for near-field thermal radiation can be found in Ref.~\cite{BiehsEtAl2021}.

In this work we propose a thermal switching effect for the HF between two hBN NPs in close vicinity to a hBN substrate. hBN is a natural hyperbolic material~\cite{Narimanov} which supports like black phosphorus or periodically ordered graphene or graphite structures directional hyperbolic surface modes~\cite{LiuEtAl2019,ShenEtAl2018,LiuXuan2016,Wu2021,Wu2021c}. Due to its intrinsic anisotropy the radiative HF between two hBN sheets can be modulated by the relative twisting of their optical axes~\cite{LiuEtAl2017,Wu2021b} an effect which is similar to the HF modulation between grating structures~\cite{BiehsEtAl2011}. In our work, we will show that the strong directionality of the energy or heat flow of the hyperbolic surface phonon polaritons (HSPP) in an hBN substrate can be used to modulate the radiative HF between two NPs by a huge amount when rotating the substrate or the NPs. We find that the modulation contrast for the full HF is about almost 1500 which is in stark contrast to the modulation contrast found between two hBN films which is about 5.36 at its maximum~\cite{LiuEtAl2017,Wu2021b} and between two gratings where it is about 10 at its maximum~\cite{BiehsEtAl2011}.   

\begin{figure}
	\centering
	\includegraphics[width=0.5\textwidth]{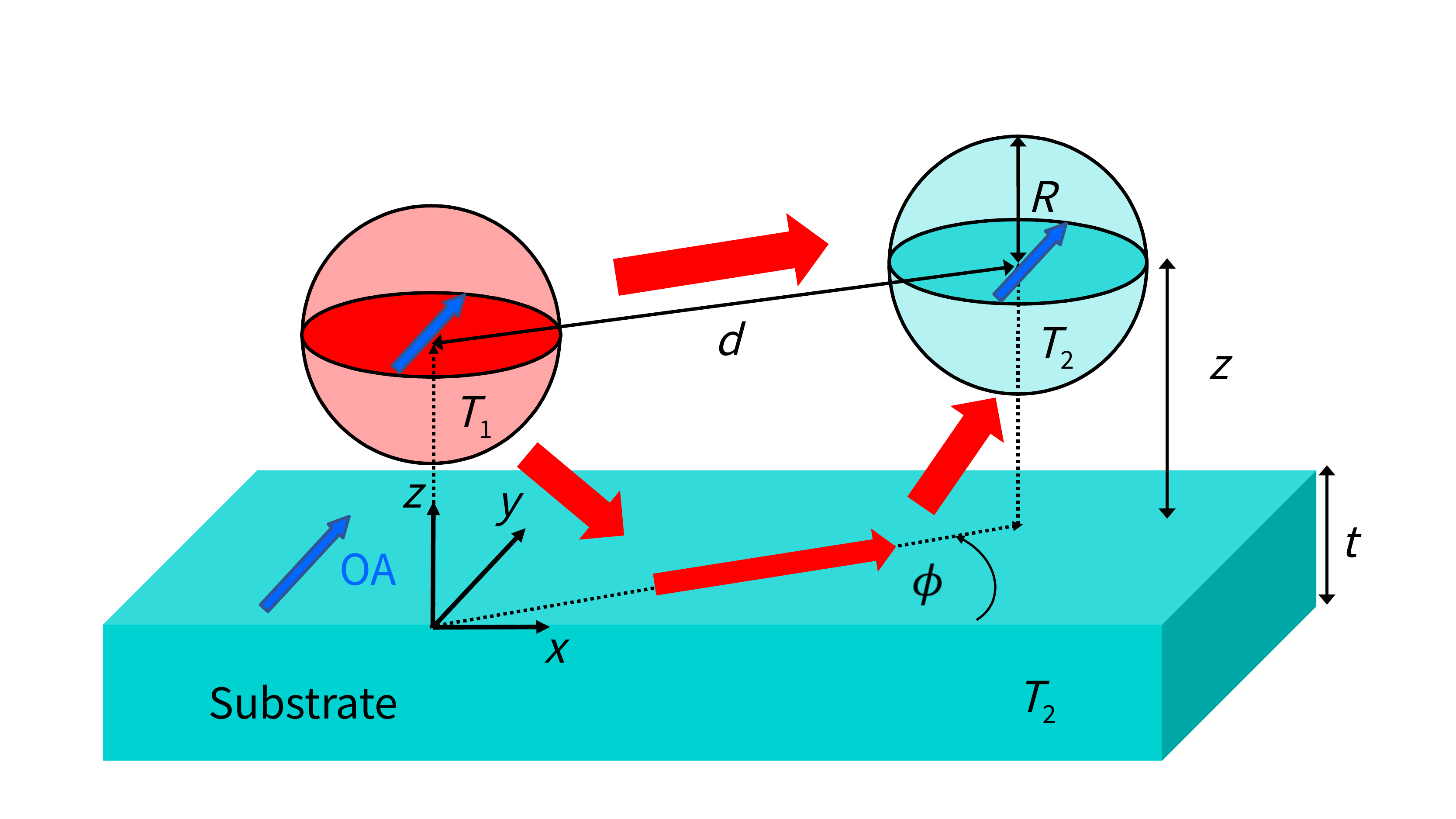}
	\caption{Sketch of the considered configuration together with a visualization of the two HF channels: direct radiative heat exchange and heat exchange via modes of the substrate like wave-guide modes and surface modes. The angle $\phi$ is the rotation angle between the optical axis and the axis connecting the two NPs.}
	\label{Fig:Geometry}
\end{figure}

%
%
\section{Setup and formalism}

We consider a setup as sketched in Fig.~\ref{Fig:Geometry} where two spherical hBN NPs (labeled with 1 and 2) are at a distance $z$ from a planar hBN substrate. The two NPs have temperatures $T_1$ and $T_2$, respectively, and the inter-particle distance is $d$. When omitting the thermal emission to the background~\cite{Domingues,Chapuis,Ott2020}, the net reveiced/emitted power for particle $i = 1,2$ can be written as ($j = 1,2; j \neq i$)
\begin{equation}
  P_i =  \int_0^\infty \!\! \frac{\rd \omega}{2 \pi}\, P_{i,\omega} = 3 \int_0^\infty \!\! \frac{\rd \omega}{2 \pi}\, \bigl[\Theta(T_j) - \Theta(T_i)\bigr] \mathcal{T}_{ji} 
\end{equation}
where $\Theta(T) = \hbar \omega / (\exp(\hbar \omega/ k_B T) - 1)$ is the Bose-Einstein function; $\hbar$
is the reduced Planck constant and $k_B$ is the Boltzmann constant. $\mathcal{T}_{ji}$ is the transmission coefficients for the power exchanged between the NPs and is given by~\cite{EkerothEtAl2017}
\begin{align}
   \mathcal{T}_{ji} &= \frac{4}{3} k_0^4 \Tr\bigl[ \uuline{\chi} \mathds{G}_{ij} \uuline{\chi} \mathds{G}_{ij}^\dagger \bigr],
\end{align}
where we have introduced the wave number in vacuum $k_0 = \omega/c$. When neglecting the radiative correction we can write the response function of the NPs $\uuline{\chi}$ in terms of the polarizability tensor $\uuline{\alpha}$ as~\cite{EkerothEtAl2017}
\begin{align}
    \uuline{\chi} = \frac{1}{2 \ri} \bigl( \uuline{\alpha} - \uuline{\alpha}^\dagger \bigr).
\end{align}
Furthermore, $\mathds{G}_{ij} = \mathds{G}(\mathbf{r}_i, \mathbf{r}_j)$ are the Green functions at positions $\mathbf{r}_{i/j}$ of particle $i$ or $j$. The detailed definition can be found in App.~\ref{App:Green}. Finally, for spherical NPs having a radius $R$ the polarizability in dipole approximation (quasi-static limit) is given by~\cite{LakhtakiaEtAl1991}
\begin{equation}
  \underline{\underline{\alpha}} = 4\pi R^3(\underline{\underline{\epsilon}}-\mathds{1})(\underline{\underline{\epsilon}}+2\mathds{1})^{-1}.
  \label{alphaerg}
\end{equation}
As a consequence, the heat exchange scales like $R^6$ in the distance regime $d,z > 4R$ where the dipole model can be applied.

%
%
\section{HVPP and HSPP in hBN}

\begin{figure*}
	\centering
	\includegraphics[width=1.0\textwidth]{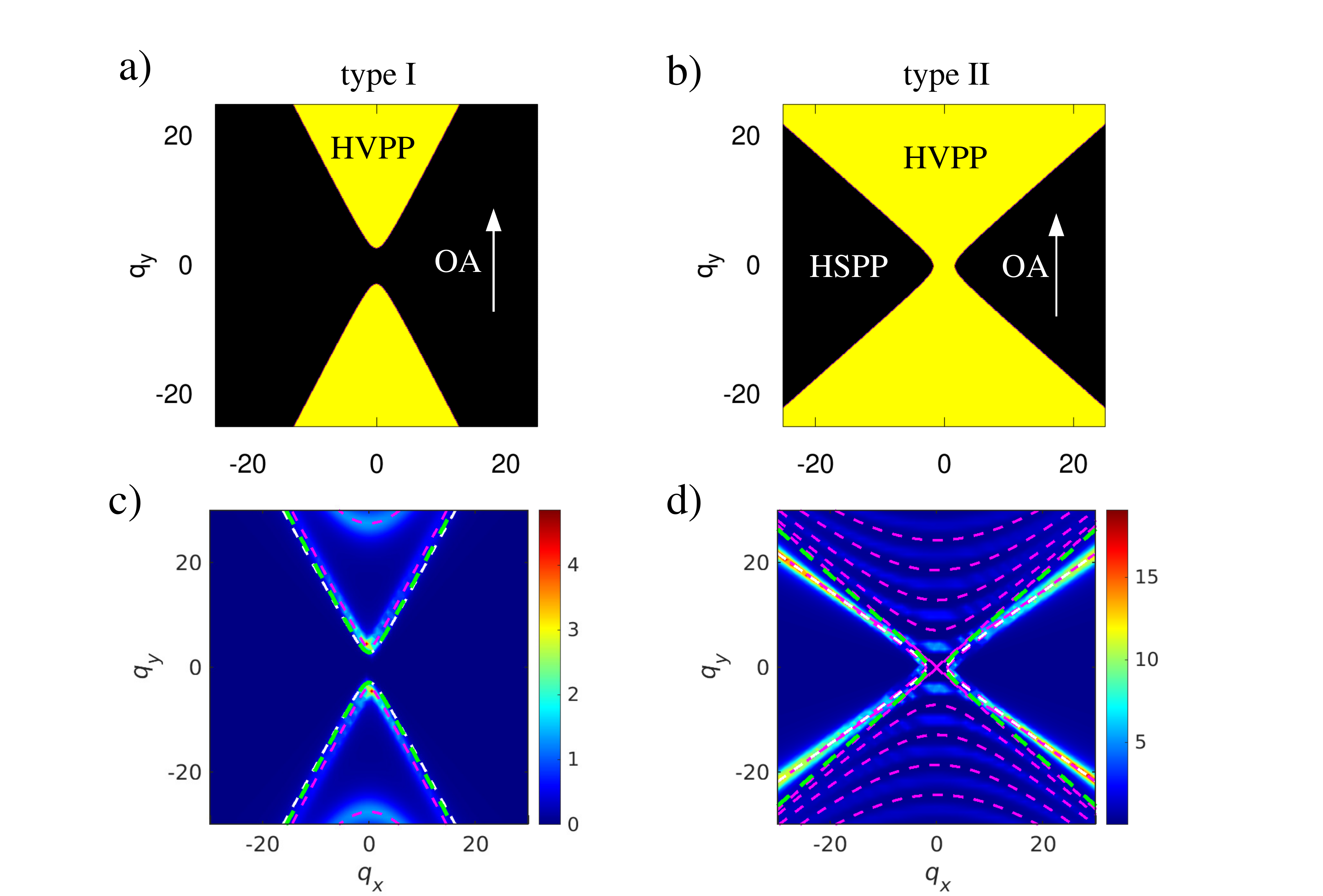}
	\caption{Regions in the $q_x$-$q_y$ plane where HVPP and HSPP can exist for a) $\omega = 1.5\times10^{14}$rad/s (type I) and b) $\omega = 2.9\times10^{14}$rad/s (type II) when the optical axis (OA) is along the y direction.  c) and d) show the corresponding $\Im(r_{pp})$. The green dashed lines are the iso-frequency lines from Eq.~(\ref{Eq:dispersionHM}) dividing the $q_x$-$q_y$ plane into the regions with HVPP and HSPP. The magenta dashed lines are the HVPP and HSPP solutions of Eq.~(\ref{Eq:dispersionSMVM}) and the white dashed lines are the solutions of Eq.~(\ref{Eq:dispersionSM}).}
	\label{Fig:ReflectionP}
\end{figure*}

In the following we assume that the NPs and the substrate are made of hBN and therefore are natural hyperbolic materials~\cite{Narimanov}. Furthermore, we assume that the optical axis is along the y direction so that the permittivity tensor can be written as
\begin{equation}
	\underline{\underline{\rm \epsilon}} = {\rm diag}(\epsilon_\perp,\epsilon_\parallel,\epsilon_\perp)
\label{epsilon}
\end{equation}
with the permittivity along/perpendicular to the optical axis~\cite{LiuXuan2016}
\begin{equation}
    \epsilon_{\perp,\parallel}=\epsilon_{\infty\perp,\parallel}\biggl(1+\frac{\omega_{L\perp,\parallel}^2-\omega_{T\perp,\parallel}^2}{\omega_{T\perp,\parallel}^2-\omega_{\perp,\parallel}^2 - \ri \gamma_{\perp,\parallel}\omega}\biggr)
\end{equation}
with $\epsilon_{\perp}=2.95$, $\omega_{L\perp}= 1610$cm$^{-1}$ , $\omega_{T\perp}= 1610$cm$^{-1}$, $\gamma_\perp= 5$cm$^{-1}$  and $\epsilon_{\perp}=4.87$, $\omega_{L\perp}=830$ cm$^{-1}$, $\omega_{T\perp}= 780$ cm$^{-1}$  and $\gamma_\perp= 4$cm$^{-1}$.  As already verified in several other works~\cite{LiuXuan2016,LiuEtAl2017,Wu2021,Wu2021b,Wu2021c} there is a hyperbolic band of type I ($\Re(\epsilon_\parallel) < 0, \Re(\epsilon_\perp) > 0$) in the frequency range $1.47-1.56\times10^{14}$ rad/s and a hyperbolic band of type II ($\Re(\epsilon_\parallel) > 0, \Re(\epsilon_\perp) < 0$) in the frequency range $2.58-3.03\times10^{14}$ rad/s. 

Due to its natural anisotropy hBN can support hyperbolic volume phonon polaritons (HVPP) as well as HSPP for the extra-ordinary modes inside the uni-axial material. The HVPP are the propagating wave solutions for waves traveling inside hBN within the hyperbolic frequency bands. When considering thin films of hBN these waves can form Fabry-P\'{e}rot-like standing waves within the thin film. On the other hand, the HSPP are evanescent wave solutions of the wave equation within the hyperbolic frequency bands which can travel within the interface but have exponentially decaying electromagnetic fields perpendicular to the interface. Therefore the HSPP are confined to the interfaces of hBN and vacuum. For thin films of hBN the HSPP of the two interfaces can couple leading to hybridized HSPPs. The $k_x$-$k_y$ range where either HVPP or HSPP can exist, can therefore be distinguished by the $z$ component of the wave vector for the considered extra-ordinary modes(eom)~\cite{LiuEtAl2017,Wu2021}, which determines whether the fields in z-direction are exponentially damped or not. In our case it is given by~\cite{LiuEtAl2017,LiEtAl2017,WuEtAl2019}
\begin{equation}
   k_{z,\rm eom} = \sqrt{k_0^2 \epsilon_\parallel - k_x^2 - k_y^2 \epsilon_\parallel/\epsilon_\perp}.
\end{equation}
When this component is purely imaginary for real-valued permittivities, then the corresponding modes would be evanescent within the uni-axial materials which are the HSPP if $k_x,k_y > k_0$. In the case that it is purely real-valued for real-valued permittivities the corresponding modes are propagating HVPP. Both regions are visualized in both hyperbolic bands in Fig.~\ref{Fig:ReflectionP}a) and b). The HVPP only exist above and below the isofrequency lines which follow from the condition $k_{z,\rm eom} = 0$, i.e. 
\begin{equation}
   \frac{k_x^2}{k_0^2 \epsilon_\parallel} + \frac{k_y^2}{k_0^2 \epsilon_\perp} = 1,
   \label{Eq:dispersionHM}
\end{equation}
whereas the HSPP can only exist on the left and right of these iso-frequency lines fulfilling the dispersion relation~\cite{LiuEtAl2017}
\begin{equation}
 \frac{k_y^2}{k_0^2} (1 - \epsilon_\parallel \epsilon_\perp) + \frac{k_x^2}{k_0^2} (1 - \epsilon_\perp^2) = - \epsilon_\parallel \epsilon_\perp . 
\label{Eq:dispersionSM}
\end{equation} 
Note that the HSPP do not exist in hBN for the type I case~\cite{Wu2021}. For thin films similar but more elaborate dispersion relations can be derived~\cite{AlvarezEtAl2019}. In the case that $k_x, k_y \gg k_0$ it reads ($\l = 0,1,2,\ldots$)
\begin{equation}
	\sqrt{k_x^2 + k_y^2} =  \frac{2 \rho}{t} \biggl[ \arctan\biggl(\frac{\rho}{\epsilon_\perp}\biggr) + \frac{\pi}{2} \l \biggr]
\label{Eq:dispersionSMVM}
\end{equation}
with $\rho = \ri \sqrt{\epsilon_\perp (k_x^2 + k_y^2)/( \epsilon_\perp k_x^2 + \epsilon_\parallel k_y^2)}$. Note, that this relation holds for HVPP and HSPP. The corresponding results are plotted in Fig.~\ref{Fig:ReflectionP}c) and d).

Due to the fact that the optical axis is within the interface, there are depolarization effects which means that an incoming s-polarized wave can be scattered into a p-polarized wave and vice versa. Therefore the reflection coefficients $r_{sp}$ and $r_{ps}$ are in general non-zero and therefore the reflection tensor 
\begin{equation}
  \mathds{R} = \begin{pmatrix} r_{ss} & r_{sp} \\ r_{ps} & r_{pp} \end{pmatrix}
\end{equation}
is not diagonal anymore. However, the near-field coupling of the two NPs via the HSPP of the substrate is mainly given by the component $r_{pp}$ of the reflection tensor describing the reflection of an incoming p-polarized wave into a p-polarized wave so that we can understand the underlying physics by focusing on that component. In Fig.~\ref{Fig:ReflectionP} we show $\Im(r_{pp})$ for an hBN substrate of thickness $t = 500$nm with optical axis in y direction for two different frequencies from the hyperbolic type I and II bands within the $q_x$-$q_y$ plane where $q_{x/y} = k_{x/y}/k_0$. Furthermore, we have added the corresponding dispersion relations of the HVPP and HSPP. Note, that the dispersion relation for semi-infinite materials coincides with the corresponding dispersion relation for a thin film when $q_x, q_y \gg 1$.

%
%
\section{Numerical results}

\begin{figure}
	\centering
	\includegraphics[width=0.5\textwidth]{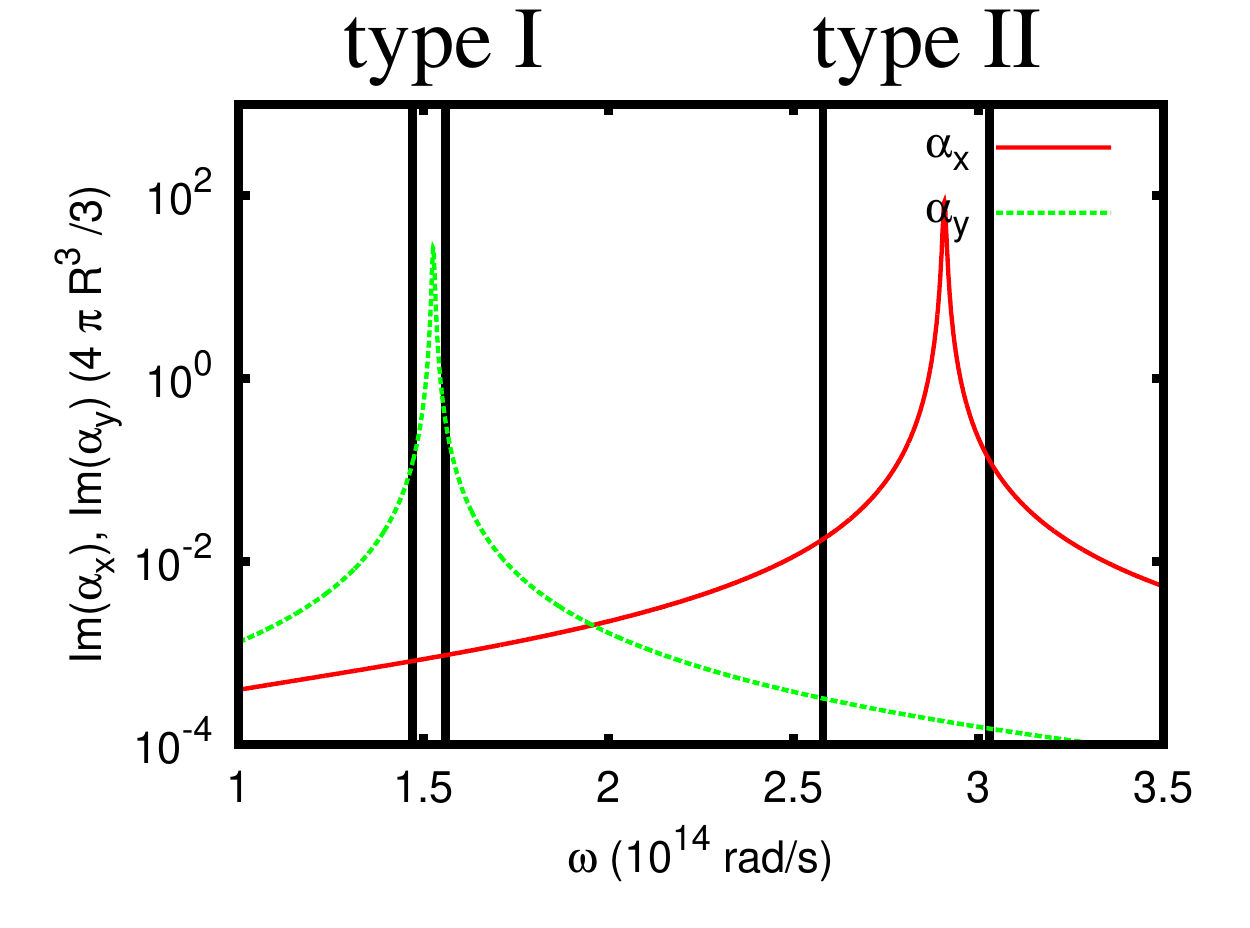}
	\caption{$\Im(\alpha_x)$ and $\Im(\alpha_y)$ for hbN NPs with optical axis in y direction normalized to the particle volume $4 \pi R^3 /3$. Note that $\Im(\alpha_z) = \Im(\alpha_x)$. The horizontal lines mark the edges of the two hyperbolic frequency bands.}
	\label{Fig:Polarizability}
\end{figure}

Now, when bringing the two hBN NPs in close vicinity to the substrate they can couple to the HVPP and HSPP and thus the heat flow between the two NPs can be tremendously enhanced, in particular, via the surface modes~\cite{Saaskilathi2014,Asheichyk2017,DongEtAl2018,paper_2sic}. In order to see which of the two hyperbolic frequency bands is more relevant for the heat transport we show in Fig.~\ref{Fig:Polarizability} the imaginary part of the components of the polarizability which is proportional to the absorptivity of the NPs. It can be observed that the component parallel to the optical axis has a resonance in the low-frequency hyperbolic band of type I and the other two components have a resonance in the high-frequency hyperbolic band of type II. This second resonance at $\omega = 2.906\times10^{14}$rad/s is much stronger than the first one so that the HF between the two NPs will be dominated by the type II band. Therefore, mainly the HSPP in this high-frequency band are responsible for the radiative coupling between the NPs.

\begin{figure}
	\centering
	\includegraphics[width=0.5\textwidth]{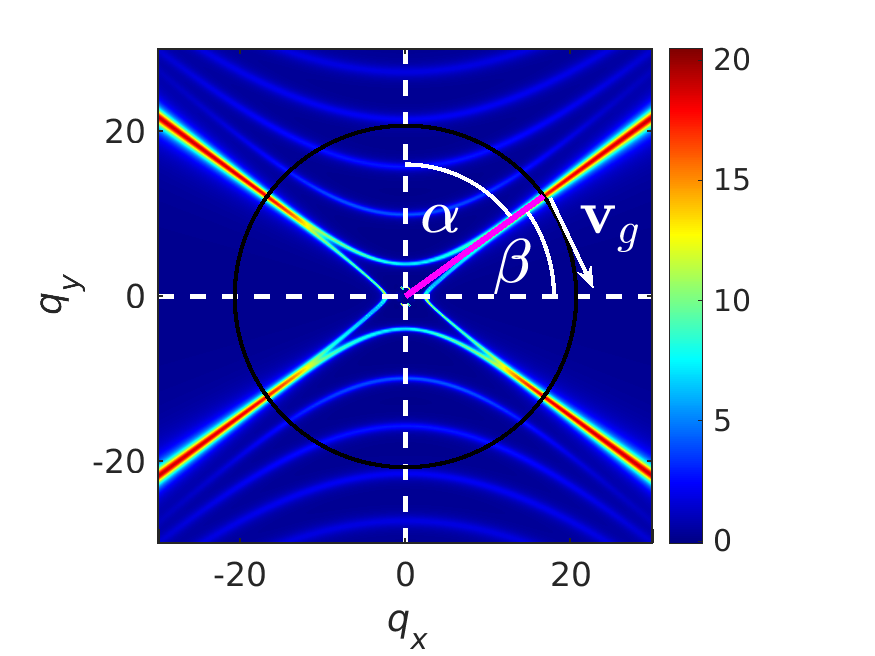}
	\caption{Imaginary part of $r_{pp}$ at $\omega = 2.9\times10^{14}\,{\rm rad/s}$ as in Fig.~\ref{Fig:ReflectionP}b) with angles $\alpha$ and $\beta$ and a sketch of the group velocity together with the asymptote (pink line) of the surface mode dispersion relation from Eq.(\ref{Eq:angles}). The circle with radius $1/(k_0 z) \approx 21$ determines the location of $(k_x,k_y)$ points which give the main contribution for the inter-particle HF via the surface at a distance of $z = 50$nm.}
	\label{Fig:DefinitionOfAngles}
\end{figure}

\begin{figure}
	\centering
	\includegraphics[width=0.4\textwidth]{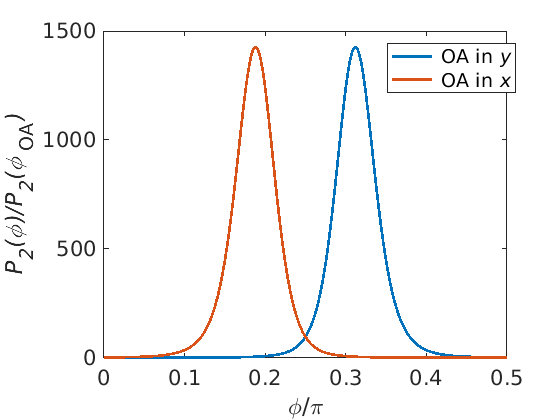}
	\includegraphics[width=0.4\textwidth]{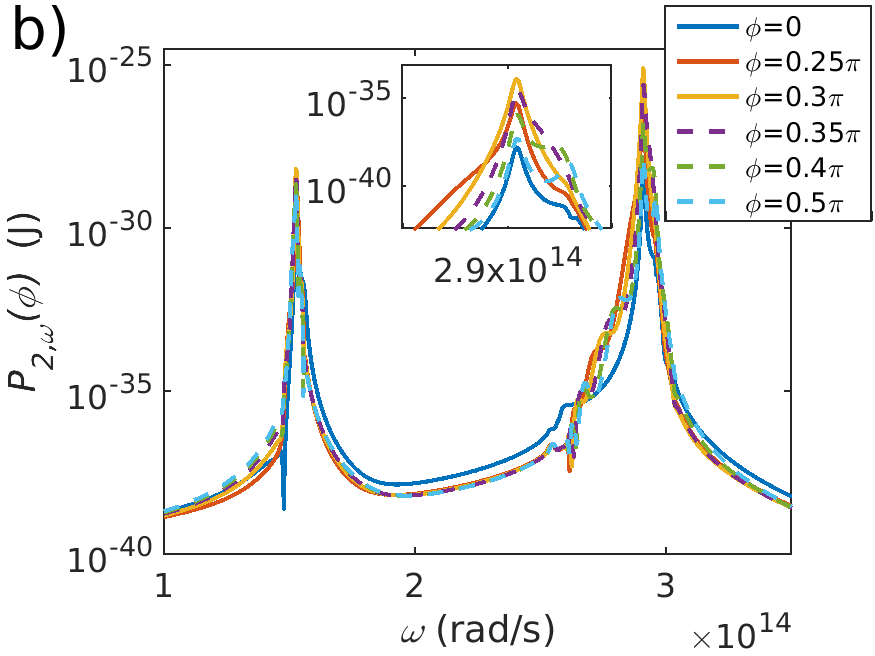}
	\caption{a) Power $P_2$ exchanged between the two NPs (with optical axis of substrate and NP either in x- or y direction) as function of the rotation angle $\phi$ of the axis connecting the NPs and the x-axis. The used parameters are $T_1 = 350$K, $T_2 = 300K$, $z = 50$nm, $t = 500$nm, $d = 700$nm and $R = 10$nm. The power is normalized to the value $P_{2} (\phi_{\rm OA})$ with $\phi_{\rm OA} = 0$ for OA in y-direction and $\phi_{\rm OA} = \pi/2$ for OA in x-direction. b) Power spectrum $P_{2,\omega}$ for the different rotational angles with OA in y-direction. Inset: enlargement of the frequency region around the resonance at $\omega = 2.906\times10^{14}\,{\rm rad/s}$.}
	\label{Fig:Rotation}
\end{figure}

Furthermore, the modes which are responsible for the heat transfer when bringing the NPs in the near-field regime of the slab are those modes for which $ \sqrt{k_x^2 + k_y^2} \approx 1/z$, because these are the modes which dominate the near-fields of the slabs~\cite{Volokitin2007,Doro2011}. Hence, when drawing a circle of radius $1/(k_0 z)$ into the $q_x$-$q_y$-plane then the modes crossing this circles will play the most important role for the inter-particle HF via the modes of the substrate as shown in Fig.~\ref{Fig:DefinitionOfAngles} where $z = 50\,{\rm nm}$ and hence $1/(k_0 z) \approx 21$. It can be seen that there are significant contributions mainly due to the large wave vector HSPP, but also by the large wave-vector HVPP which have a crossing with the circle at the angles $\beta$ or $\alpha$, resp. Hence, when rotating the substrate or the NPs in the x-y plane one can expect to have a large inter-particle HF when the inter-particle axis is aligned to the direction of the group velocity 
\begin{equation}
  \mathbf{v}_g = \nabla_\mathbf{k} \omega = \begin{pmatrix} \frac{\rd \omega}{\rd k_x} \\ \frac{\rd \omega}{\rd k_y} \end{pmatrix} = \begin{pmatrix} \bigl( \frac{\rd k_x}{\rd \omega} \bigr)^{-1} \\ \bigl(\frac{\rd k_y}{\rd \omega}\bigr)^{-1} \end{pmatrix}
\end{equation}
of the HSPP which can be evaluated from the dispersion relation of the HSPP. As depicted in Fig.~\ref{Fig:DefinitionOfAngles} the group velocity is always perpendicular to the isofrequency lines. Therefore we can easily determine its direction in the following approximative manner. In the long-wave vector limit $q_x \gg 1$ and $q_y \gg 1$ the HSPP for a film converge to the bulk dispersion relation in Eq.~(\ref{Eq:dispersionSM}) which can be approximated in this limit by
\begin{equation}
	\frac{k_y}{k_x} = \frac{q_y}{q_x} \approx \sqrt{\frac{\epsilon_\perp^2 - 1}{1 - \epsilon_\parallel \epsilon_\perp}} = \tan(\beta)
\label{Eq:angles}
\end{equation}
where $\beta$ is the angle between the optical axis and the large wave-vector branch of the surface mode. If now the axis connecting the NP is along the y-axis, then we have to rotate the substrate or the NPs by the angle $\beta$ to align the inter-particle axis with the direction of the group velocity. On the other hand, if the NPs are aligned along the x-axis it is necessary to rotate the substrate or the NPs by the angle $\alpha$ where $\tan(\alpha) \tan(\beta) = 1$. Since, at  $\omega = 2.906\times10^{14}$rad/s we have $\Re(\epsilon_\perp) = -2$ and $\Re(\epsilon_\parallel) = 2.8$ we find $\beta = 0.19\pi$ and $\alpha = 0.31\pi$. To verify this reasoning we show in Fig.~\ref{Fig:Rotation}a) the power $P_{2}$ as a function of the rotation angle $\phi$ of the substrate when starting with both NPs aligned along the x-axis (and for two orientations of the optical axis). It can be seen that there is indeed the maximum at the expected position which is more than 1400 times larger than the for transmitted power when the axis connecting the NPs is perpendicular to the OA. Hence, we find a huge modulation contrast due to the strong directionality of the HSPP. Furthermore, in Fig.~\ref{Fig:Rotation}b) we show that for all rotational angles the high-frequency resonance dominates the power flow between the NPs in this configuration. Note, that the direction of the rotation is due to the symmetry irrelevant.

\begin{figure}
	\centering
	\includegraphics[width=0.4\textwidth]{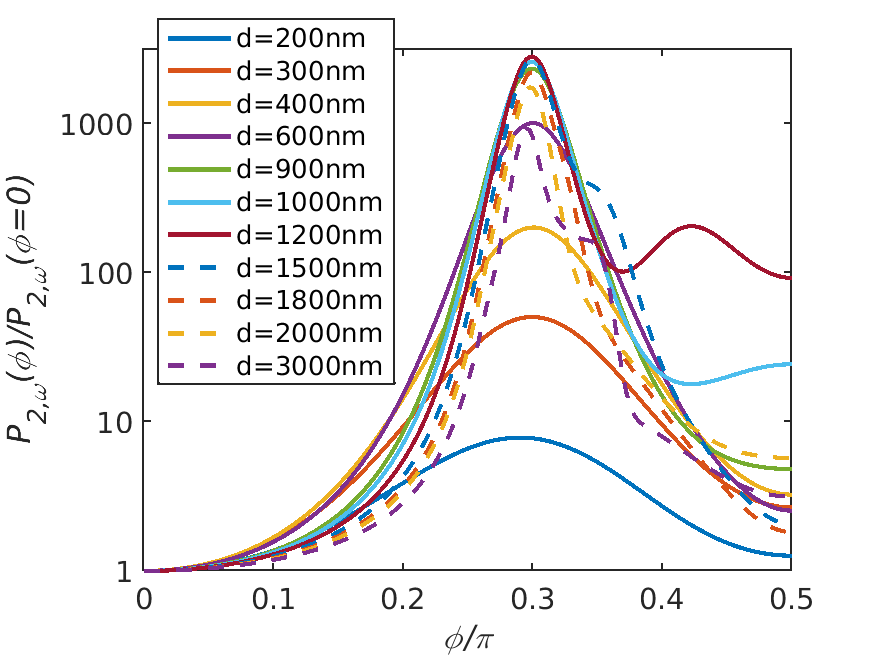}
	\caption{Normalized spectral power $P_{2,\omega}(\phi)/P_{2,\omega}(\phi = 0)$ as a function of the rotation angle $\phi$ for different inter-particle distances $d$ using parameters as in Fig.~(\ref{Fig:Rotation}) and optical axis of substrate and NPs in y direction.}
	\label{Fig:Rotation2diffd}
\end{figure}

\begin{figure}
	\centering
	\includegraphics[width=0.5\textwidth]{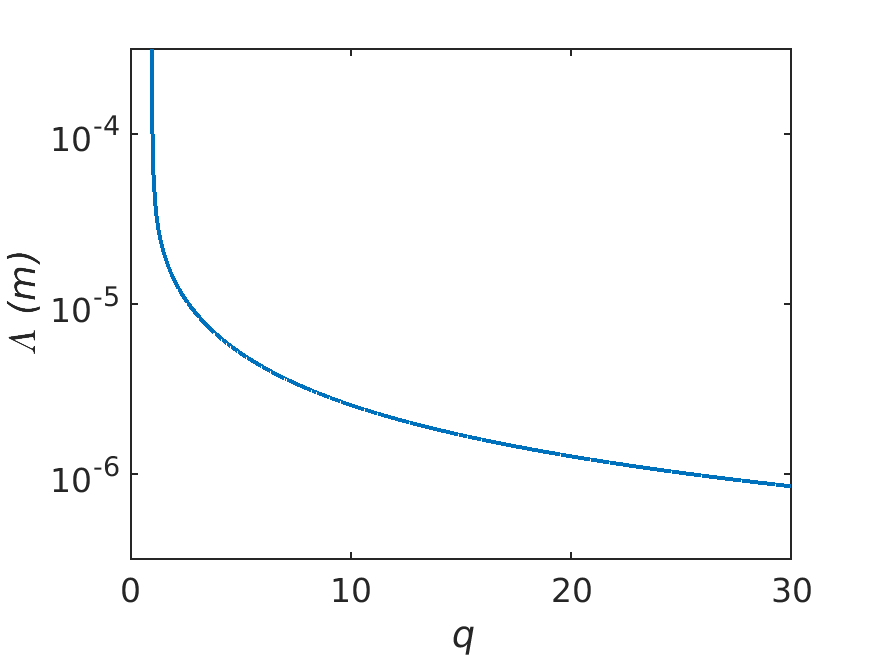}
	\caption{Propagation length $\Lambda$ of the HSPP at $\omega = 2.9\times10^{14}$rad/s from Eq.~(\ref{Eq:Propagationlength}) for a semi-infinite substrate as function of $q = \sqrt{q_x^2 + q_y^2}$.}
	\label{Fig:propagationlength}
\end{figure}

\begin{figure}
	\centering
	\includegraphics[width=0.4\textwidth]{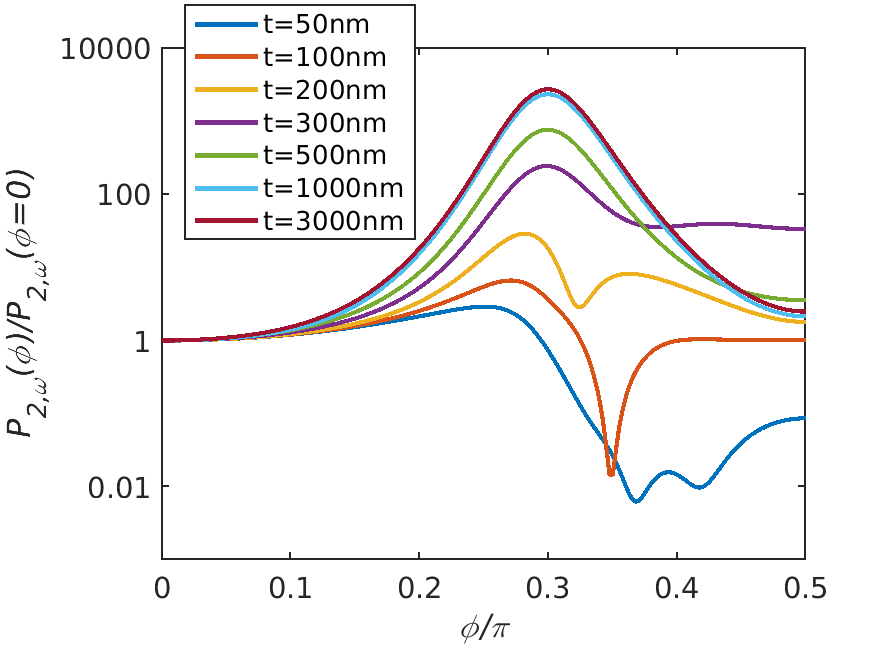}
	\caption{As in Fig.~\ref{Fig:Rotation2diffd} for $d = 700$nm and varying $t$.}
	\label{Fig:Rotation2difft}
\end{figure}

In Fig.~\ref{Fig:Rotation2diffd} we show how the modulation of the transferred power $P_{2,\omega}$ at the resonance frequency $\omega = 2.9\times10^{14}$ for different rotational angles $\phi$ depends on the inter-particle distance $d$ when starting with NPs axis along x-direction ( $\phi = 0$) and optical axis along y direction. First, it is obvious that there is always a maximum at $\alpha \approx 0.31 \pi$ as expected from our explanation. Furthermore, it is clear that for angles larger than the asymptote of the dispersion relation of the HVPP with the y-axis also HVPP contribute to the HF so that at $\phi = \pi/2$ the inter-particle HF is in general larger than for $\phi = 0$. It can further be observed that there is an optimal distance of about $d = 1200$nm at which the enhancement at $\phi = 0.31\pi$ has its maximum of about 3500 times $P_{2,\omega} (\phi = 0)$. Note that of course for $d \ll z$ the particle-particle heat transfer is just a direct heat transfer via the vacuum. By increasing the inter-particle distance the HVPP and HSPP in the substrate will more and more contribute to the inter-particle HF until the distance $d$ is getting larger than the propagation length of the HSPP or HVPP. Then the HF via between the NPs will decay when further increasing the inter-particle distance. 

\begin{figure*}
	\centering
	\includegraphics[width=0.4\textwidth]{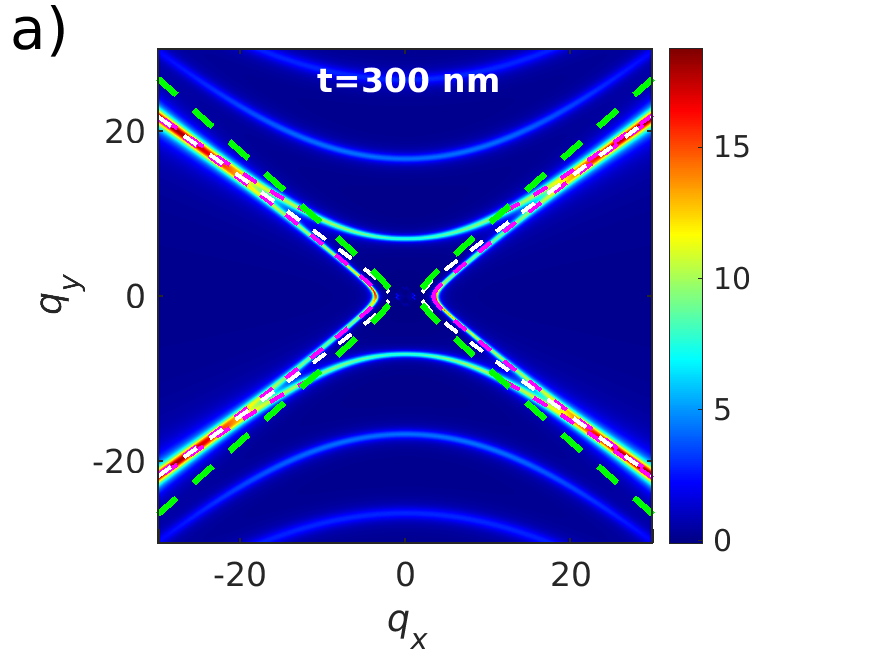}
	\includegraphics[width=0.4\textwidth]{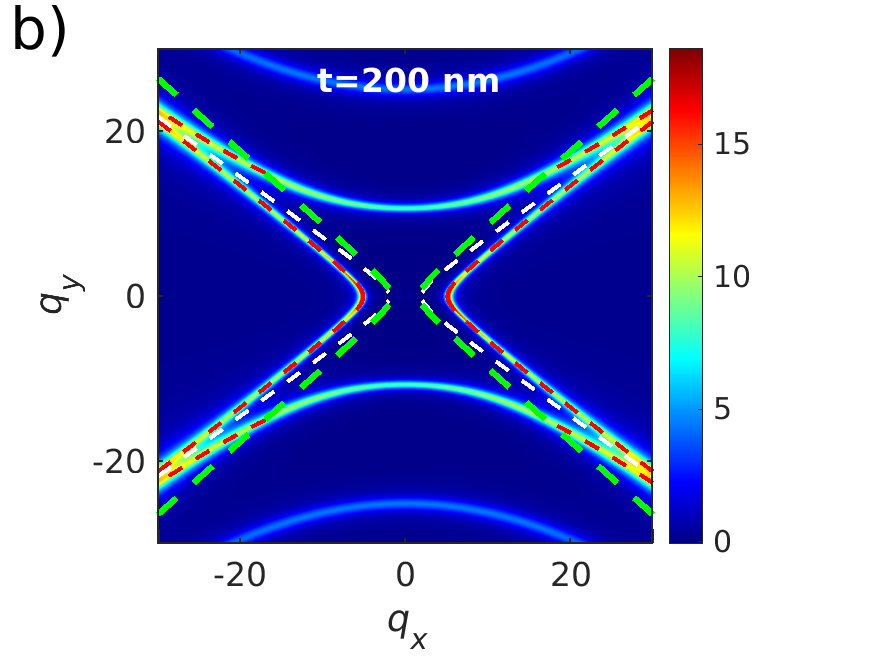}\\
	\includegraphics[width=0.4\textwidth]{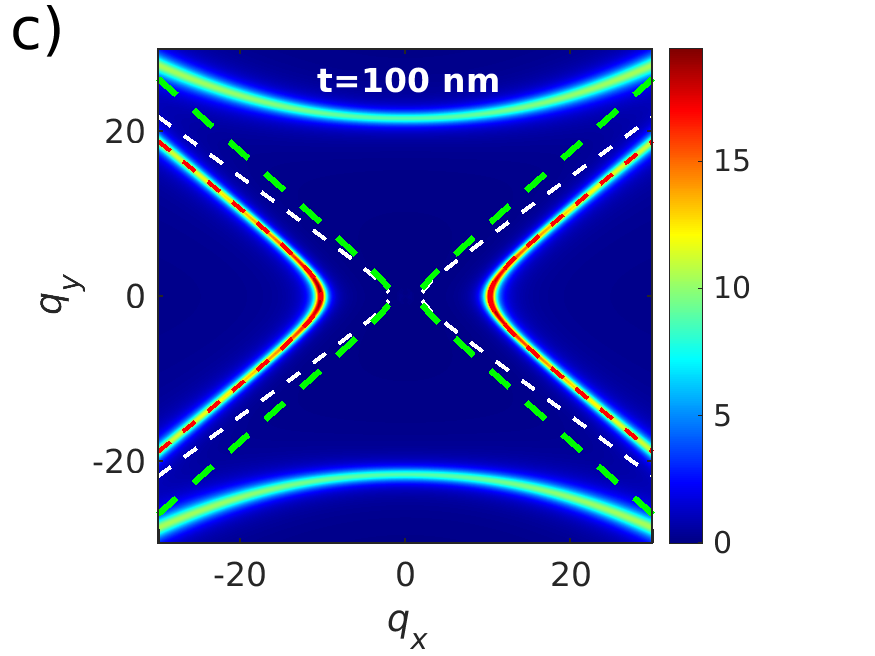}
	\includegraphics[width=0.4\textwidth]{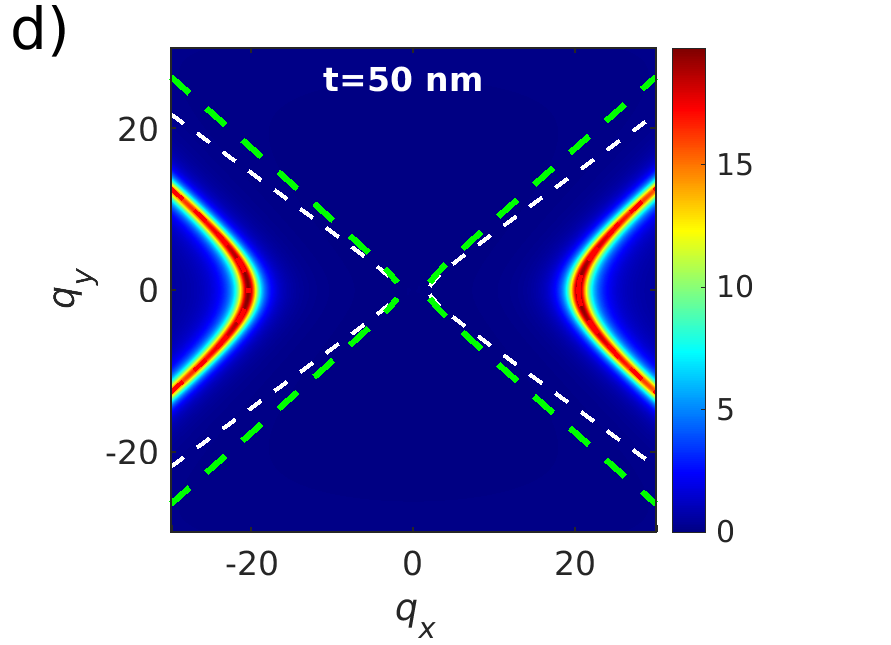}
	\caption{Imaginary part of $r_{\rm pp}$ for different thicknesses of the substrate at $\omega = 2.9\times10^{14}$rad/s: a) $t = 300$nm, b) $t = 200$nm, c) $t = 100$nm, d) $t = 50$nm. The dashed lines have the same meaning as in Fig.~\ref{Fig:ReflectionP}.  }
	\label{Fig:imagrppdifft}
\end{figure*}

To see that the distance $d$ of maximal heat transfer is intimately connected to the propagation length $\Lambda$ of the HSPP we first define it by
\begin{equation}
  \Lambda = - \frac{|\mathbf{v}_{\rm g}|}{2 \omega''}
	\label{Eq:Propagationlength}
\end{equation}
where $\omega''$ is the imaginary part of the complex solution of the dispersion relation in Eq.~(\ref{Eq:dispersionSM}) or (\ref{Eq:dispersionSMVM}), resp., assuming real $k_x$ and $k_y$. Since the fields of the HSPP have a phase $\exp(\ri k_x x + \ri k_y y - \ri \omega t)$, $\omega''$ is negative and the damping constant for the intensity is $-1/2 \omega''$. In Fig.~\ref{Fig:propagationlength} we have plotted the propagation length of the HSPP at $\omega = 2.9\times10^{14}\,{\rm rad/s}$. Obviously, at $q = 1/(k_0 z) \approx 21$ we have a value of $1.5\,\mu{\rm m}$ which is very good agreement with the maximum found at  $d = 1200$nm. Hence, by increasing the distance $z$ of the nanoparticles to the substrate the maximal HF increase shifts to larger $d$ and vice versa.

Finally, in Fig.~\ref{Fig:Rotation2difft} we also show the impact of the substrate thickness $t$ on the spectral transferred power at $\omega = 2.9\times10^{14}$rad/s. It can be observed that the maximum angle is shifted to smaller angles when making the substrate thinner. For thicknesses smaller than 300nm clearly a second maximum at a larger angle can be observed. To understand this behaviour we show in Fig.~\ref{Fig:imagrppdifft} the corresponding plots of $\Im(r_{\rm pp})$ which clearly show the mode structure available for the inter-particle HF. It can be seen that the surface and propagating mode dispersion are pushed to larger wave vectors when making the film thinner. Due to this effect the dispersion of the HSPP in the first quadrant is getting slightly steeper so that the angle $\beta$ is slightly increased and hence $\alpha$ is slightly decreased. This explains the shift of the first maximum in Fig.~\ref{Fig:imagrppdifft}. That the maximum also drops to smaller value when making the substrate thinner can also be attributed to the fact that less and less HSPP can contribute since the dispersion relation is pushed out of the circle with radius $21$. The contributions for angles larger than the maximum are again due to the HVPP. As for the HSPP for thin layers less and less of these modes can contribute to the radiative HF explaining the decreasing inter-particle HF for thicknesses below 300nm and angles larger than the maximum angle. For very thin substrates of only $t = 50$nm these modes do simply not exist in the wave vector region responsible for the heat transfer as can be seen in Fig.~\ref{Fig:imagrppdifft}d) explaining the very small HF for angles larger than the maximum angle for $t = 50\,{\rm nm}$. It can also be understood that there is a sudden HF drop for thin films for angles slightly larger than the maximum angle. This drop falls into the angle region where we have no HSPP and no HVPP. When making the films thinner this region becomes larger and can at a certain thickness directly be observed in the angle dependence of the inter-particle HF. Hence, we can associate the second maximum to the contribution of the propagating HVPP. These are of course also the reason for the maxima at larger angles than $0.31\pi$ in Fig.~\ref{Fig:Rotation2diffd}.

%
%
\section{conclusion}

In summary, we have shown that the radiative HF between two NPs in close vicinity to a natural hyperbolic material like hBN can be highly modulated by exploiting their coupling to the highly directional HSPP. The modulation contrast can be almost 1500 when rotating the axis connecting the NPs with respect to the optical axis of the substrate. This contrast is especially high for optically thick films. For thin films with a thickness $t$ on the order of the distance $z$ or smaller the HF between the NPs cannot couple anymore to the HSPP so that in this case the HF is efficiently inhibited. Apart from its potential usefulness for thermal management at the nanoscale the observed effect might also be employed for a future multitip HF measurement~\cite{PBA2019} of the HSPP dispersion relation itself, since it is highly dependent on the dispersion relation of the HSPP. A direct measurement of the modulation effect might be possible with a many-body heat transfer setup like that recently developed by Reddy's group~\cite{Reddy}.

%
%
\section{Acknowledgment}

S.-A.\ B. acknowledges support from Heisenberg Programme of the Deutsche Forschungsgemeinschaft (DFG, German Research Foundation) under the project No. 404073166. Xiaohu Wu acknowledges support from Natural Science Foundation of Shandong Province under the project No. ZR2020LLZ004. Yang Hu acknowledges support from Key Research and Development Program of Shaanxi Province under the project No. 2018SF-387 and Basic Research Program of Taicang under the project No. TC2019JC01.

\appendix

\section{Alternative derivation of the angles for maximal heat flux}

An alternative route to the foregoing around Eq.~(\ref{Eq:angles}) is to determine first the components of the group velocity
\begin{align}
  \frac{\rd k_x}{\rd \omega} &= k_y \frac{\rd}{\rd \omega} \sqrt{\frac{1 - \epsilon_\parallel \epsilon_\perp}{\epsilon_\perp^2 - 1} } =  k_y \frac{\rd}{\rd \omega}  \tan(\alpha), \\
  \frac{\rd k_y}{\rd \omega} &= k_x \frac{\rd}{\rd \omega} \sqrt{\frac{\epsilon_\perp^2 - 1}{1 - \epsilon_\parallel \epsilon_\perp}  } = k_x \frac{\rd}{\rd \omega}  \tan(\beta)
\end{align}
so that the group velocity is
\begin{equation}
    \mathbf{v}_g =  \begin{pmatrix} \frac{1}{k_y \frac{\rd}{\rd \omega}  \tan(\alpha)} \\ \frac{1}{k_x \frac{\rd}{\rd \omega}  \tan(\beta)} \end{pmatrix}.
\end{equation}
Therefore, the angle of the group velocity with the y-axis is determined by 
\begin{equation}
\begin{split}
  \tan(\tilde{\beta}) &= \frac{ |k_x \frac{\rd}{\rd \omega}  \tan(\beta)| }{| k_y \frac{\rd}{\rd \omega}  \tan(\alpha)|} \\
                       &= \tan(\alpha)  \biggl| \frac{\frac{\rd}{\rd \omega}  \tan(\beta) }{\frac{\rd}{\rd \omega}  \tan(\alpha)} \biggr| \\
                       &=  \tan(\alpha)  \biggl| - \frac{\frac{\rd \beta}{\rd \omega}  \frac{1}{\sin^2(\beta)} }{\frac{\rd \beta}{\rd \omega}  \frac{1}{\cos^2(\beta)}} \biggr|  \\
                       &= \tan(\alpha) \tan^2(\beta) \\
                       &= \tan(\beta).
\end{split}
\end{equation}
Obviously, we obtain the same result as with the graphical method.

\section{Green Function} \label{App:Green}

The Green function $\mathds{G}_{ij} = \mathds{G}_{ij}^0 + \mathds{G}_{ij}^{\rm sc}$ consists of a vacuum part $\mathds{G}_{ij}^0$ and a scattering part $ \mathds{G}_{ij}^{\rm sc}$. The vacuum part $\mathds{G}^0_{ij}$ is given by~\cite{Novotny} 
\begin{equation}
   \mathds{G}^{0}_{ij} = \frac{e^{\ri k_0{d}}}{4\pi d} (a\mathds{1}+b\mathbf{e}_{ij}\otimes\mathbf{e}_{ij})
\end{equation}
with $d = |\mathbf{r}_i - \mathbf{r}_j|$ and
\begin{equation}
  a = 1+\frac{ik_0{d}-1}{k_0^2d^2}, b =\frac{3-3ik_0{d}-k_0^2d^2}{k_0^2d^2}, 
\end{equation}
and
\begin{equation}
  \mathbf{e}_{ij} = \frac{\mathbf{r}_{i}-\mathbf{r}_{j}}{d}.
\end{equation}

The scattering part $\mathds{G}_{ij}^{\rm sc}$ is the contribution due to the presence of the film. Assuming that $z_i = z_j = z$ it can be expressed within the Weyl representation as~\cite{paper_2sic,Ott2019,Ott2020}
\begin{equation}
  \mathds{G}^{sc}_{ij} = \int\frac{{\rm d}^2 \kappa}{(2\pi)^2}e^{\ri \boldsymbol{\kappa} \cdot(\mathbf{x}_i-\mathbf{x}_j)}\frac{\ri \re^{2 \ri \gamma_0 z}}{2\gamma_0} \!\! \sum_{\alpha,\beta=p,s} \!\! r_{\alpha\beta}\mathbf{a}_\alpha^{\,+}\otimes\mathbf{a}_\beta^{\,-}
\end{equation}
with $\gamma_0 =\sqrt{k_0^2-\kappa^2}$, $\boldsymbol{\kappa}= (k_x, k_y)$, $\mathbf{x}_{i/j}=(x_{i/j},y_{i/j})^t$, and the two polarization vectors
\begin{equation}
  \mathbf{a}_s^\pm = \frac{1}{\kappa}\begin{pmatrix}	k_y\\-k_x\\0	\end{pmatrix} \quad\text{and} \quad
  \mathbf{a}_p^\pm = \frac{1}{\kappa k_0} \begin{pmatrix}\mp	k_x\gamma_0\\\mp k_y\gamma_0\\\kappa^2	\end{pmatrix}.
\end{equation}
The double integral over the whole range of $k_x$ and $k_y$ cannot be further simplified because due to the anisotropy of the considered material hBN the reflection coefficients $r_{ss}$, $r_{sp}$, $r_{ps}$, and $r_{pp}$ depend on $\boldsymbol{\kappa}$ and not only $\kappa$ as for isotropic materials. We evaluate these integrals numerically taking as input the reflection coefficients of the hBN film which are derived with the method detailed in App.~\ref{appenB}.
Note, that there is a typo in the expression for the scattering part of the Green function in Refs.~\cite{Ott2019,Ott2020}: there the factor 2 in the exponential appears as a prefactor.

\section{Reflection coefficients} \label{appenB}

Here, we sketch the derivation of the reflection coefficients for a uni-axial film with optical axis in a plane parallel to the film interfaces. In the following the two interfaces of the film of thickness $t$ are assumed to be at $z=0$ and $z = -t$. Taking the translational symmetry parallel to the film interfaces into account the electromagnetic field within the film can be expressed by
\begin{eqnarray}
  \mathbf{E}_m= \mathbf{S}(z)e^{\ri(k_xx + k_yy - \omega t)} \\
  \mathbf{H}_m=\sqrt{\frac{\epsilon_0}{\mu_0}}\mathbf{U}(z)e^{\ri(k_xx + k_yy - \omega t)}
\end{eqnarray}
with $\mathbf{S}=(S_x,S_y,S_z)^t$ and $\mathbf{U}=(U_x,U_y,U_z)^t$ which are for the moment unknown. For further simplification we define $K_y \equiv k_y/k_0$, $K_x \equiv k_x/k_0$ and $z'=zk_0$. Now, applying the macroscopic Maxwell equations
\begin{align}
  \nabla\times\mathbf{E}_m = \ri \omega \mu_0 \mathbf{H}_m, \\
  \nabla\times\mathbf{H}_m = - \ri \omega \epsilon_0 \uuline{\epsilon} \mathbf{E}_m,
\end{align}
for a uni-axial medium (with optical axis in x-y plane) described by the permittivity tensor $\uuline{\epsilon}$, we obtain
\begin{equation}
  \partial_{z'}\begin{pmatrix} S_x\\S_y\\U_x\\U_y \end{pmatrix} 
      = \mathds{A} \begin{pmatrix} S_x\\S_y\\U_x\\U_y \end{pmatrix}
\label{diff}
\end{equation}
with 
\begin{equation}
{\small  \mathds{A} = \ri \begin{pmatrix}
  0 & 0& \frac{ K_x K_y}{\epsilon_{zz}}&1 -\frac{K_x^2}{\epsilon_{zz}} \\
  0 & 0& -1  + \frac{K_y^2}{\epsilon_{zz}}&- \frac{K_xK_y}{\epsilon_{zz}} \\
  - K_xK_y -  \epsilon_{yx} &  K_x^2 - \epsilon_{yy}&0&0\\
   \epsilon_{xx} - K_y^2 &  K_xK_y + \epsilon_{xy}&0&0
  \end{pmatrix}.}
\end{equation}
The solution of Eq.~(\ref{diff}) can be expressed by linear combinations
{\small
\begin{align}
  S_x(z)=\sum_{m=1}^{2}w_{1,m}c_m e^{k_0q_m(z-z_t)}+\sum_{m=1}^{2}w_{1,2+m}c_m e^{k_0q_{m+2}z} \\
  S_y(z)=\sum_{m=1}^{2}w_{2,m}c_m e^{k_0q_m(z-z_t)}+\sum_{m=1}^{2}w_{2,2+m}c_m e^{k_0q_{m+2}z} \\
  U_x(z)=\sum_{m=1}^{2}w_{3,m}c_m e^{k_0q_m(z-z_t)}+\sum_{m=1}^{2}w_{3,2+m}c_m e^{k_0q_{m+2}z} \\
  U_y(z)=\sum_{m=1}^{2}w_{4,m}c_m e^{k_0q_m(z-z_t)}+\sum_{m=1}^{2}w_{4,2+m}c_m e^{k_0q_{m+2}z}
\end{align}
}
of the four eigenvectors $\mathbf{w}_m = (w_{1,m}, w_{2,m}, w_{3,m},q_{4,m})^t$ ($m = 1,2,3,4$) of $\mathds{A}$ to the four eigenvalues 
$q_m$ with $\Im(q_{m}) > 0$ ($m = 1,2$) and $\Im(q_{m}) < 0$ ($m = 3,4$). The four coefficients $c_m$ are fixed by imposing the boundary conditions.

To impose the boundary conditions we need also an ansatz for the fields outside the film. Let's start with the incident and reflected fields in the region $z > 0$. We make the ansatz
\begin{align}
  \mathbf{E}_i &= \bigl( A_s \mathbf{a}_s^- + A_p \mathbf{a}_p^- \bigr) \re^{\ri (k_x x + k_y y - k_z z - \omega t )}, \\
  \mathbf{H}_i &=  \sqrt{\frac{\epsilon_0}{\mu_0}} \bigl( - A_s \mathbf{a}_p^- + A_p \mathbf{a}_s^- \bigr) \re^{\ri (k_x x + k_y y - k_z z-\omega t )} 
\end{align}
and
\begin{align}
  \mathbf{E}_r &= \bigl( R_s \mathbf{a}_s^+ + R_p \mathbf{a}_p^+ \bigr) \re^{\ri (k_x x + k_y y + k_z z - \omega t )}, \\
  \mathbf{H}_r &= \sqrt{\frac{\epsilon_0}{\mu_0}} \bigl( - R_s \mathbf{a}_p^+ + R_p \mathbf{a}_s^+ \bigr) \re^{\ri (k_x x + k_y y + k_z z - \omega t )} 
\end{align}
with $k_z = \sqrt{k_0^2 - \kappa^2}$. Similarly, we can make the ansatz for the transmitted fields in region $z < -t$
\begin{align}
  \mathbf{E}_t &= \bigl( T_s \mathbf{a}_s^- + T_p \mathbf{a}_p^- \bigr) \re^{\ri (k_x x + k_y y - k_z z - \omega t )}, \\
  \mathbf{H}_t &=  \sqrt{\frac{\epsilon_0}{\mu_0}} \bigl( - T_s \mathbf{a}_p^- + T_p \mathbf{a}_s^- \bigr) \re^{\ri (k_x x + k_y y - k_z z - \omega t )}. 
\end{align}
Finally, by demanding that the field components parallel to the interface are continuous at $z = 0$ and $z = -t$ (which are the boundary conditions), we obtain eight equations which fix the unknown eight coefficients $c_m$, $R_{s/p}$, and $T_{s,p}$ for given $A_s$ and $A_p$. In particular, the reflection coefficients are determined by $r_{ss} = R_s/A_s$, $r_{ps} = R_p/A_s$ for an arbitrary $A_s \neq 0$ and $A_p = 0$  and $r_{sp} = R_s/A_p$, $r_{pp} = R_p/A_p$ for an arbitrary $A_p \neq 0$ and $A_s = 0$.

\color{black}

\end{document}